\documentclass[a4paper,12pt]{article}

\newcommand{\sect}[1]{\setcounter{equation}{0}\section{#1}}

\newcommand{\bea}{\begin{eqnarray}}
\newcommand{\ena}{\end{eqnarray}}

\renewcommand{\a}{\alpha}
\renewcommand{\b}{\beta}

\newcommand{\e}{\epsilon}
\newcommand{\dsl}{\pa \kern-0.5em /}

\newcommand{\pa}{\partial}

\begin{document}
\topmargin 0pt
\oddsidemargin 0mm

\renewcommand{\thefootnote}{\fnsymbol{footnote}}
\begin{titlepage}
\begin{flushright}
MCTP-01-01\\
hep-th/0102056
\end{flushright}

\begin{center}
{\Large \bf  (1 +
 p)-Dimensional Open D(p - 2) Brane Theories}
\vspace{5mm}
\begin{center}
{\large
        J. X. Lu\footnote{jxlu@umich.edu}
\\}
\vspace{5mm}

{\em Michigan Center for Theoretical Physics, Randall Physics
 Laboratory\\
 University of Michigan, Ann Arbor
 MI 48109-1120, USA\\}
\end{center}
\end{center}

\vspace{5mm}
\centerline{{\bf{Abstract}}}
\vspace{5mm}
The dynamics of a Dp brane can be described either by an
open string ending on this brane or by an open D(p - 2) brane ending on 
the same Dp brane. The ends of the open string couple to a Dp brane 
worldvolume gauge field while the boundary of the open D(p - 2) brane 
couples to a (p - 2)-form worldvolume potential whose field strength 
is Poincare dual to that of the gauge field on the Dp-brane
worldvolume. With this in mind, we find that the Poincare dual of the 
fixed rank-2 magnetic field used in defining a (1 + p)-dimensional 
noncommutative Yang-Mills (NCYM) gives precisely a near-critical
electric field for the open D(p - 2) brane. We therefore find 
(1 + p)-dimensional open D(p - 2) brane theories along the same line as 
for obtaining noncommutative open string theories (NCOS), OM
theory and open Dp brane theories (ODp) from NS5 brane. Similarly, the 
Poincare dual of the near-critical electric field used in defining
 a (1 + p)-dimensional NCOS gives a fixed magnetic-like field. This
field along with the same bulk field scalings defines a (1 +
p)-dimensional noncommutative field theory. In the same spirit, we can
have various (1 + 5)-dimensional noncommutative field theories resulting 
from the existence of ODp if the description of open D(4 - p) brane
ending on the NS5 brane is insisted. 

\end{titlepage}

\newpage
\renewcommand{\thefootnote}{\arabic{footnote}}
\setcounter{footnote}{0}
\setcounter{page}{2}

\sect{Introduction}
By now we know that there exists not only the big M-theory but also a 
little m-theory. The latter is particularly interesting since it 
shares many properties of the big M-theory and yet appears as a
decoupled theory without gravity.  Therefore, we have a better
hand on this theory and hopefully we can learn new things and
gain insights for the big M-theory in the process of uncovering more on this 
little m-theory.

	The purpose of this paper is to show the existence of
new non-gravitational theories which are closely related to the recently 
discovered decoupled noncommutative Yang-Mills theories 
(NCYM) \cite{wu,conds,douh,seiw}, 
noncommutative open string theories (NCOS) \cite{seist,gopmms,harone}, 
 OM theory and open Dp brane (ODp) theories\cite{gopmss,berbss,hartwo}. 

In particular, we will show that the Dp brane worldvolume Poincare dual
of the fixed rank-2
magnetic field used in defining a (1 + p)-dimensional NCYM 
gives a critical (p - 1)-form electric field\footnote{The
scalings for the bulk fields such as the metric and the closed string
coupling remain the same.} for an
open D(p - 2) brane ending on the Dp brane. We therefore find 
(1 + p)-dimensional open D(p - 2) brane theories in the same spirit as
for OM theory, NCOS and ODp.
In other words, with the same bulk (the metric and the closed string
coupling) scaling limit, we can end up with either
 a (1 + p)-dimensional open D(p - 2) brane theory from 
the open D(p - 2) brane perspective or a (1 + p)-dimensional
NCYM from the open string perspective. Moreover, the open D(p - 2) brane theory provides a
completion of the NCYM if the latter is nonrenormalizable. In this
sense, the former is in general a better description.

By the same token, we find that the existence of a (1 + p)-dimensional
NCOS implies also a (1 + p)-dimensional ``noncommutative'' field 
theory\footnote{Here for $p > 3$, the noncommutative geometry is also 
expected to be nonassociative as well.}.
 The corresponding noncommutative  geometry is
determined through the quantization of the
boundary action which is obtained from a topological one for 
the open D(p - 2) brane.
For the particular $p = 3$ case, the new noncommutative field theory is
also a NCYM resulting from an open  D-string
ending on a D3 brane and can actually be identified with the usual 
NCYM resulting from an open F-string ending on the same base D3 brane.

The above results are consistent with the compactification of
 OM theory on either a magnetic circle or an electric circle.
The usual picture is: the compactification of OM theory on a magnetic
 circle gives the usual (1 + 4)-dimensional NCYM while on an electric
circle it gives the (1 + 4)-dimensional NCOS. As we will show in section
 5, the actual path is: The magnetic-circle compactification of OM
 theory gives our (1 + 4)-dimensional open D2 brane theory 
 which  provides a completion of the effective (1 + 4)-dimensional
 NCYM. The electric-circle compactification gives the (1 +
 4)-dimensional NCOS which provides a completion of the new effective  (1 +
 4)-dimensional noncommutative tensor field theory mentioned above. We will
elaborate  these in section 5.

Along the similar line, we should also have new (1 + 5)-dimensional
noncommutative field
theories given the existence of the ODp theories from NS5 brane for $p \le 5$.
We will discuss this in section 6. All these new
non-gravitational theories are  consistent with U-duality, therefore
lending support to the notion that U-duality is inherited to the little
m-theory without gravity.

This paper is organized as follows: In section 2, we give a rather
detailed motivation for the work presented in this paper. In section 3,
we show that the fixed rank-2 magnetic field used in defining 
 a usual (1 + p)-dimensional NCYM from the open string perspective 
gives precisely a critical (p - 1)-form electric field 
 for an open D(p - 2) brane theory if the dynamics of 
the base Dp brane is described in terms of the ending D(p - 2) brane. 
We also discuss the relationship between
the open D(p -2) brane theory and the corresponding NCYM.
 In section 4, we follow the same line as 
in section 3 but now for a (1 + p)-dimensional NCOS. We will show that the 
resulting limit gives a noncommutative field theory
 with a noncommutative 
geometry determined by the boundary action for the D(p - 2) brane. 
In section 5, we give a detailed picture on the compactification of OM
theory on either a magnetic or an electric circle.
We will show that the results obtained in the previous sections are
consistent with the compactifications of OM theory.
 In section 6, we first argue the proper limits for ODp theories from
NS5 brane. Then we show that the (1 + p)-dimensional open
D(p - 2) brane theories discussed in section 3 
are U-duality related to the ODp's. We also show that the 
bulk decoupling limits for ODp from NS5 brane give ones for noncommutative
field theories living on NS5 brane in a similar spirit as discussed in
section 3 and 4.  In section 7, we
discuss S-duality between the (1 + 3)-dimensional NCOS and our open
D-string theory, and the implication for the existence of a (1 +
3)-dimensional open (p, q) string theory. 

\sect{Motivation}
	Strominger some time ago in \cite{str} concluded that a D(p - 2)
brane can end on a Dp brane (also M2 brane on M5 brane) 
without violating charge conservation along the similar line for a
fundamental string on a Dp brane. This same conclusion was also reached
by Townsend in \cite{tow} from the analysis of Chern-Simons terms in 
$D = 10$ and $D = 11$ supergravity theories. From the D-brane
worldvolume perspective, the end of a fundamental string (or F-string)
 appears as
a point electric charge which couples to the worldvolume U(1)
field. The magnetic charge (or monopole) with respect to the U(1) field 
 implies actually a (p - 2)-dimensional extended object carrying an 
electric-like charge which couples to a worldvolume (p - 1)-form field
strength (Poincare
dual to the U(1) gauge field strength)  in the 
Poincare-dual picture. Therefore, a (1 + p)-dimensional NCYM as a
decoupled theory of 
 Dp-branes with a magnetic field in the F-string picture implies the
existence of a different decoupled theory of the Dp brane in an
electric-like (p - 1)-form field strength in the open D(p - 2) brane
picture. This new theory is just our (1 + p)-dimensional open D(p - 2)
brane theory which will be discussed in the following section.
 Similarly, a (1 + p)-dimensional NCOS as a decoupled theory of 
Dp brane with a near-critical electric field in the F-string picture
implies also the existence of a different field theory of Dp brane
with a magnetic-like (p - 1)-form field strength in the open D(p - 2)
brane picture. This new theory is a ``noncommutative'' field theory defined on
a noncommutative geometry.

	Let us elaborate the above further. The dynamics of Dp-brane
with a constant magnetic flux in it  can be described by the open
F-string ending on the
Dp-brane with its boundary coupled to this background. In the decoupling
limit, the kinetic term of the string theory can be ignored and
the dynamics is described by a topological term~\cite{seiw}.  This
topological term can be expressed as a boundary one and the quantization
of this boundary action gives rise to spatial noncommutativity along
the directions with nonvanishing magnetic field on the Dp brane
worldvolume. 

What is the picture if we look from the description in terms of  
the open D(p - 2) brane ending on the Dp-brane with the same scalings for
the bulk metric and the closed string coupling as those for
NCYM?  As is well known that Dp
branes with a constant magnetic flux 
represent a non-threshold bound state of Dp branes with smeared D(p - 2)
branes along the two co-dimensions\cite{bremm,rust,cosp}. The smeared D(p -
2) branes are within the Dp-brane worldvolume rather than end on them.
As discussed in \cite{lur}, in the decoupling limit for NCYM, if we
view the smeared D(p - 2) branes as periodic vortices along the two 
co-dimensions, each vortex will decouple from the rest. Therefore we
need to consider only one vortex, for example, the one in the origin of
the coordinate system for the two co-dimensions. In other words, we have
localized D(p - 2) branes within the Dp brane worldvolume in the
decoupling limit for NCYM. We now know that in terms of the open D(p - 2)
brane picture,  this system should also decouple from the bulk in the
decoupling limit and
its dynamics is described  by the open D(p - 2) branes which couple to  a 
Dp-brane
worldvolume (p - 1)-form field strength. The very fact that the D(p - 2) 
branes reside within Dp brane worldvolume must imply that the background 
(p - 1)-form electric field reach its critical 
value\footnote{This conclusion can
only be drawn in the
decoupling limit. From NCYM side, we know that in the decoupling limit
the open string massive modes decouple and the dynamics is described 
by its massless modes, i.e., the gauge modes, which live on the brane.
 So we expect that the dynamical degrees of freedom should also remain
on the brane if the open D(p - 2) brane description is adopted. Here
what left in the decoupling
limit is the D(p - 2) branes and therefore the background field must
reach its critical value.}.  We will show that this is indeed true as 
expected.   

	The above picture is along the same line as for the decoupling
limits for NCOS, OM theory and those ODp from NS-5 branes. In particular,
 the gravity systems used for their gravity descriptions
\cite{gopmss,harone,lurone,bers,hartwo} in the
respective decoupling limits   are
nothing but the corresponding non-threshold bound states. For example, 
for OM theory, the gravity system is the (M5, M2) bound state\cite{towone}.
For NCOS, the gravity systems are the (F, Dp) bound state\cite{lurtwo}.
The gravity description of the present open D(p - 2) brane theory is the
same as the corresponding one of the usual (1 + p)-dimensional NCYM 
except that we have traded the asymptotic B-field for NCYM with the 
asymptotic RR (p - 1)-form potential through the Dp-brane worldvolume Poincare
duality\footnote{We will use the constant bulk B-field or RR (p -
1)-potential only when we discuss the gravity dual descriptions. 
Otherwise, we always use the worldvolume fields to avoid possible
confusions.}.

	We have the following two additional pieces of evidence to
support the existence of the
open D(p - 2) brane theories found in this paper. First, 
OM theory results from a critical electric 3-form $H_{012}$ 
field limit.
The non-linear self-duality constraint for this 3-form field implies also
a non-vanishing $H_{345}$. As discussed in \cite{gopmss}, this theory
 reduces to a usual (1 + 4)-dimensional NCYM upon compactification on a
magnetic circle. The $H_{345}$ gives a rank-2 magnetic field which gives
rise to the noncommutativity in the
NCYM theory.

	Upon the reduction on a circle along one of the M5 worldvolume 
directions, the 3-form field strength
on M5 brane will give either a 2-form gauge field strength or a 3-form field 
strength but not both on the D4 brane worldvolume. Otherwise, we 
double counting the degrees of freedom for the worldvolume field 
since the two are not independent
but related through a constraint inherited from the self-duality on M5 
brane. This is familiar for the self-dual 5-form field strength
in the dimensional reduction of type IIB supergravity on a circle to
the $N = 2$ nine dimensional supergravity.

The usual (1 + 4)-dimensional NCYM is
nonrenormalizable and therefore this description is  an effective one 
which is good for relevant energy much smaller than 
 the inverse of the gauge coupling $g^2_{\rm
NCYM}$. If this effective description is valid, 
we can choose to keep the 2-form gauge field
strength rather than 3-form field strength. 
 
Note that the magnetic-circle compactification of OM theory
 is along a direction
transverse to the open membrane which is used to define OM theory. 
One must be wondering where is the open membrane and naturally expects
an open membrane theory in (1 + 4)-dimensions.
In other words, we expect OM theory to reduce to an open membrane theory
in (1 + 4)-dimensions when the compactification radius is invisible to
the OM theory (i.e., the KK modes are too heavy in comparison to the OM
theory scale). This theory is also expected to provide a complete
description in (1 + 4)-dimension. As we will show in section 5, 
this is indeed true.
This open membrane theory is just our (1 + 4)-dimensional open D2 brane 
theory which we will discuss in the following section. 
This theory provides the completion of
the usual (1 + 4)-dimensional NCYM. In other words, OM theory implies
the existence of the (1 + 4)-dimensional open D2 brane theory. 
For this theory, we need to keep instead the 3-form $H_{012}$ upon the reduction.
 Starting with this (1 + 4)-dimensional open D2 brane theory,
we can obtain in general (1 + p)-dimensional open D(p - 2) brane
theories by T-duality along a direction either common or transverse to both
of D(p - 2) and Dp branes. We limit ourselves to $p \le 5$ in this paper
because for $p > 5$, the corresponding (1 + p) NCYM cannot decouple from
the bulk\cite{aliio}. This might imply that we have only decoupled open Dp 
brane theories for $p \le 3$.

	By the same token, we may expect a new noncommutative 
 tensor field theory upon the compactification of OM theory
on an electric circle when the spatial 3-form $H_{345}$ can be kept
instead. We
will discuss this possibility in section 4.    	 

	The ODp theories from NS-5 brane discovered in \cite{gopmss,hartwo}
also imply the existence of the (1 + p)-dimensional open D(p - 2)
brane theories found in this paper. 
As discussed in \cite{gopmss}, one direct evidence for ODp theories
is from the fact that an open string ending on a D5 brane is S-dual to
a D-string ending on a NS-5 brane in type IIB string theory. The former gives
the (1 + 5)-dimensional NCOS in the critical electric field limit.
The S-dual of this gives OD1 now also in the corresponding 
critical electric field
limit. This can also be understood as the electric force, due to 
the near-critical  electric
field, acting at the two ends of the D-string on the NS
-5 brane almost balances the D-string tension. As a result, the D-string
decouples from the bulk and is confined on the NS-5 brane worldvolume.
T-dualities along NS5 brane directions on this OD1 give in general ODp
for $p \le 5$. In other words, these ODp are just the results of 
open Dp branes ending on the base NS5 in the corresponding critical
electric field limits.

	The direct connection between these ODp and the ones found in
this paper occurs for OD3.
Since the tension and the near-critical electric field associated with 
the open D3 brane, and the scalings for the closed string
parameters (metric and closed string coupling) remain the same under
S-duality, we 
 conclude that  the S-dual of OD3 gives another OD3 since the D3 brane
itself is intact under S-duality\footnote{Some parameters of the original
OD3 theory such as the effective open D3 brane coupling are transformed
under S-duality but the theory is not. This conclusion differs from that
given in \cite{gopmss} where the S-duality gives (1 + 5)-dimensional
NCYM. We will reconcile this difference in section 6.}. 
This new OD3 theory is now from an open D3 brane ending
on D5 branes in the critical 4-form electric field
limit. Therefore, this OD3 theory is our present (1 + 5)-dimensional
open D3 brane theory. T-dualities along the D3 brane directions
therefore give also our (1 + p)-dimensional open D(p - 2) brane
theories.

	The field theories resulting from the existence of NCOS or ODp
can be discussed in a similar fashion and we will not repeat them here.

\sect{(1 + p)-Dimensional Open D(p - 2) Brane Theories}

In this section, we will show that the decoupling limit for a (1 +
p)-dimensional NCYM with
rank-2 noncommutative matrix from the open string perspective
 gives precisely a critical field limit for an open D(p - 2) brane
theory if this open D(p - 2) brane description of Dp-brane is insisted.
Let us begin with a summary of the decoupling limit for NCYM \cite{seiw}:
\begin{eqnarray}
&& \tilde \a' = {\e}^{1/2} \tilde \a'_{\rm eff}, \ \ 
\tilde g_s = \frac{\tilde \a'^{\frac{3 - p}{2}}_{\rm eff} \tilde g^2_{\rm NCYM}}{(2 \pi)^{p - 2}}
\e^{\frac{3 - (p - 2)}{4}},\ \ g_{\mu\nu} = \eta_{\mu\nu}\ (\mu, \nu =
0, 1, \cdots (p - 2)),\nonumber\\
&&g_{ij} = \e \delta_{ij}, \ (i, j = (p - 1), p),\ \ g_{mn} = \e
\delta_{mn}~(m,n = {\rm transverse}), \nonumber\\
&& 2\pi\tilde \a' B_{(p - 1) p} = \e^{1/2}, 
\label{eq:ncymdl}
\end{eqnarray}
where $\tilde g^2_{\rm NCYM}$ is the fixed noncommutative
 Yang-Mills coupling. We know that with
the presence of Dp brane, the worldvolume gauge invariant quantity is
${\cal F} = 2\pi \tilde \a' (B + F)$ with $F$ the worldvolume gauge field. For
the purpose of performing the worldvolume Poincare duality in the
following, we replace the constant rank-2 B-field in
Eq. (\ref{eq:ncymdl})
by a constant rank-2 gauge field strength using a gauge choice. As a
result, we have now
\begin{equation}
2\pi\tilde \a' F_{(p - 1) p} = \e^{1/2}, \ \ B = 0.
\label{eq:magf}
\end{equation}

The Dp-brane worldvolume Poincare dual of the above magnetic background 
gives an electric-like worldvolume (p - 1)-form field strength 
$H_{012\cdots (p - 2)}$ which is associated with the D(p - 2) brane
ending on the Dp-brane.
Note that the relevant Dp-brane Lagrangian for the purpose of obtaining
such an electric-like background field $H_{012\cdots (p - 2)}$ is
\begin{equation}
{\cal L}_{DBI} = - \frac{1}{(2 \pi)^p \tilde \a'^{(1 + p)/2} \tilde g_s} \sqrt{- \det
(g_{\a\b} + 2\pi \tilde \a' F_{\a\b})},
\end{equation}
where $\a, \b = 0, 1, \cdots p$.
We then have
\begin{equation}
\sqrt{- \det g} \frac{H_{012\cdots (p - 2)}}{2\pi} = - \frac{1}{2}
\frac{\e_{012\cdots (p - 2) ij}}{\sqrt{- \det g}} 
\frac{\partial {\cal L}_{DBI}}{\partial F_{ij}},
\label{eq:hdf}
\end{equation}
where we define $\e_{\a_0 \cdots \a_p} = g_{\a_0 \b_0} \cdots
g_{\a_p \b_p} \e^{\b_0 \cdots \b_p}$ with $\e^{01\cdots p} = 1$.

Using the scalings for $\tilde g_s$, the metric in Eq. (\ref{eq:ncymdl}) and
the magnetic background in Eq. (\ref{eq:magf}), we have from the above
\begin{equation}
H_{012\cdots (p - 2)} = \frac{1}{(2\pi)^{p - 2} \tilde \a'^{\frac{(p - 2) +
1}{2}}_{\rm eff} \tilde G^2_{o(p -2)}} \left(\frac{1}{\e} - \frac{1}{2}\right),
\label{eq:ncf}
\end{equation}
where we have defined
\begin{equation}
\tilde G^2_{o(p - 2)} = \frac{\tilde g^2_{\rm NCYM} \tilde \a'^{(3 - p)/2}_{\rm eff}}{(2\pi)^{p - 2}}.
\label{eq:odpc}
\end{equation}
The scalings for the metric and the closed string coupling remain the
same as those given in Eq. (\ref{eq:ncymdl}). The form of the above
electric (p - 1)-form field strength indicates that it reaches its
critical limit as $\e \to 0$. Let us confirm this. The effective action
of an open D(p - 2) brane ending on a Dp brane can be written in its 
simplest form as
\begin{equation}
S_{(p - 2)} = - \frac{1}{(2 \pi)^{p - 2} \tilde \a'^{(p - 1)/2} \tilde g_s}\int_{M^{p
- 1}} d^{p - 1} \sigma \sqrt{- \det(\hat g_{\mu\nu} + 2\pi\tilde \a' F_{\mu\nu})} + 
\int_{M^{p - 1}} {\cal H}_{p - 1} + \cdots,
\label{eq:dp2a}
\end{equation}
where we have 
\begin{equation}
{\cal H}_{p - 1} = C_{p - 1} + H_{p - 1},
\end{equation}
 with $C_{p - 1}$ the pull-back of the bulk RR (p - 1)-form potential
and $H_{p - 1}$
is the aforementioned Dp brane worldvolume $(p - 1)$-form field strength
which comes from the conversion of the open D(p - 2) brane boundary term
to its worldvolume along the Dp-brane directions. The $\cdots$ terms are 
irrelevant for the discussion of this paper and for this reason we
 drop them from now on. The D(p - 2) brane worldvolume gauge field 
$F_{\mu\nu}$ is also irrelevant and we drop it for the following
discussion. In the above, the gauge invariant quantity is 
now ${\cal H}_{p - 1}$. Once again, we see that in the presence of 
this D(p - 2) brane, given ${\cal H}_{p - 1}$ and $H_{p - 1}$, $C_{p -
1}$ cannot be arbitrary but fixed according to the above 
equation\footnote{This example indicates that we cannot choose the
 asymptotic values as we wish for bulk potentials whether they are NSNS or RR
origins in the presence of various kinds of D branes}. 
For the choice of Eq. (\ref{eq:ncf}), we have $C_{01\cdots (p - 2)} = 0$.

With the above, let us calculate the effective proper (also coordinate) 
tension for a D(p
- 2) brane along $12\cdots (p - 2)$ directions with the metric and
the closed string coupling given 
in Eq. (\ref{eq:ncymdl}) and with the  $H_{01\cdots (p - 2)}$ given
in Eq. (\ref{eq:ncf}), we then have
\begin{equation}
- \frac{1}{(2\pi)^{p - 2} \tilde \a'^{(p - 1)/2} \tilde g_s} + 
\e^{01\cdots (p - 2)} H_{01\cdots (p - 2)} = - 
\frac{1}{2 (2\pi)^{p - 2} \tilde \a'^{(p - 1)/2}_{\rm eff} \tilde G^2_{o(p - 2)}},
\end{equation}
which indicates that our $H_{01\cdots (p - 2)}$ is indeed a
near-critical electric field. The near-critical electric force stretches
the boundary of the D(p - 2) brane to balance its original tension such
that a finite tension as given above\footnote{As always, the resultant
finite tension is smaller than the expected one by half. For examples, 
we have this for NCOS and (1 + 5)-dimensional ODp from NS5-brane. 
This implies that
the usual evaluation of such tensions may not be completely correct.
We try to resolve this in~\cite{lu}.} is obtained. As a result, the D(p -
2) brane is now confined within the Dp-brane worldvolume. The
conventional discussion implies that we end up with an open D(p -2) brane
theory for $p \le 5$. For later use, let us summarize the decoupling 
limit for a (1 + p)-dimensional open D(p - 2) brane theory:
\begin{eqnarray}
&&\tilde \a' = \e^{1/2} \tilde \a'_{\rm eff}, \ \
\tilde g^{(p - 2)}_s = 
\e^{\frac{3 - (p - 2)}{4}} \tilde G^2_{o(p - 2)},\ \ g_{\mu\nu} = 
\eta_{\mu\nu}\ (\mu, \nu = 0, 1, \cdots (p - 2)),\nonumber\\
&&g_{ij} = \e \delta_{ij}, \ (i, j = (p - 1), p), \ \ g_{mn} = \e
\delta_{mn}~(m, n = {\rm transverse}), \nonumber\\
&&H_{012\cdots (p - 2)} = \frac{1}{(2\pi)^{p - 2} 
\tilde \a'^{\frac{(p - 2) + 1}{2}}_{\rm eff}
\tilde G^2_{o(p -2)}} \left(\frac{1}{\e} - \frac{1}{2}\right),
\label{eq:odpdl}
\end{eqnarray}
where the coupling $\tilde G_{o(p - 2)}$ 
for the open D(p - 2) brane theory is related to the
gauge coupling through (\ref{eq:odpc}). 

Let us briefly discuss each of the open D(p - 2) brane theories for $ 2 \le p
\le 5$.

\noindent
{\bf Open D0 theory}: This case can be discussed similarly following that for
 the OD0 theory from NS5 brane given in~\cite{gopmss}. The present open
 D0 brane theory results from a D2 brane in the presence of
 a worldvolume near-critical 1-form field strength 
$H_0 = \frac{1}{\e \sqrt{\tilde \a'_{\rm eff}}  \tilde G^2_{o(0)}} (1 -
 \frac{\e}{2})$. This field strength can be traded to a 1-form bulk RR
 potential $C_0$. The dynamical objects in this theory are the light D0
 branes. Again, the light excitations of this open D0 brane theory carry
 a conserved charge.

If we lift this open D0 brane theory to eleven dimensions on a
transverse circle, the D2 brane
now becomes an M2 brane. We have the eleven-dimensional Planck mass 
and the compactified radius  as
\begin{equation}
R_{11} = \sqrt{\tilde \a'} \tilde g^{(0)}_s = \e \sqrt{\tilde \a'_{\rm eff}}
\tilde G^2_{o(0)} \equiv \e R, \ \ M_p = \frac{1}{\sqrt{\tilde \a'}
(g^{(0)}_s)^{1/3}} = \e^{- 1/2} \tilde M_{\rm eff},
\end{equation}
where $\tilde M_{\rm eff} = \frac{1}{\sqrt{\tilde \a'_{\rm eff}}
\tilde G^{2/3}_{o(0)}}$.  
Choosing the fixed coordinate in the 11-th direction such that $x^{11}
\sim x^{11} + 2\pi R$, the bulk 11-dimensional metric is
\begin{equation}
d s^2_M = - (d x^0)^2 + R^2_{11} (\frac{d x^{11}}{R} - C_0 d x^0)^2 + \e
d x^2_{\perp} = \e \left[- (d x^0)^2 - d x^{11} d x^0 + d
x^2_{\perp}\right],
\end{equation}
where we have dropped a term proportional to $\e^2$. Note that the
lifted theory is defined with respect to the metric $d s^2_M/\e$ and now
the compactified 11-th direction is light-like. We have now the bulk
Planck scale $\tilde M_{\rm eff}$ which is the same as the proper
tension for the open D0 brane theory. 

	In other words, the open D0 brane theory with N units of D0
brane charge is a DLCQ compactification of M theory with N units
of DLCQ momentum in the presence of a transverse M2 brane.

\noindent
{\bf Open D1 theory}: The decoupling limit for this theory can be
summarized as
\begin{eqnarray}
&&\tilde \a' = \e^{1/2} \tilde \a'_{\rm eff}, \ \
\tilde g^{(1)}_s = 
\e^{\frac{1}{2}}\tilde G^2_{o(1)},\ \ g_{\mu\nu} = 
\eta_{\mu\nu}\ (\mu, \nu = 0, 1),\nonumber\\
&&g_{ij} = \e \delta_{ij}, \ (i, j = 3, 4), \ \ g_{mn} = \e
\delta_{mn}~(m, n = {\rm transverse}), \nonumber\\
&&H_{01} = \frac{1}{(2\pi) 
\tilde \a'_{\rm eff}
\tilde G^2_{o(1)}} \left(\frac{1}{\e} - \frac{1}{2}\right).
\label{eq:od1dl}
\end{eqnarray}
For this particular case, given the relation between the open D-string
and the open F-string, we expect that the open D-string metric and
noncommutative parameter can be obtained from the usual Seiberg-Witten
relations for open F-string ending on a D-brane through the following 
replacements:
\begin{equation}
\tilde \a' \to \tilde \a'\tilde g^{(1)}_s, \ \ \tilde g^{(1)}_s \to
\frac{1}{\tilde g^{(1)}_s}, \ \
F_{\a\b} \to H_{\a\b},
\end{equation}
i.e., we have now
\begin{eqnarray}
&&G_{\a\b} = g_{\a\b} - (2\pi \tilde \a' \tilde g^{(1)}_s)^2 (H g^{-1} H)_{\a\b},\nonumber\\    
&&\Theta^{\a\b} = 2\pi \tilde \a' \tilde g^{(1)}_s \left(\frac{1}{g +
2\pi\tilde \a' \tilde g^{(1)}_s
H}\right)^{\a\b}_A,
\end{eqnarray}
where $A$ in $()_A$ denotes the anti-symmetric part of the matrix and 
$\a,\b = 0, 1, 2, 3$. Using the above scalings, we have the open
D-string metric and the nonvanishing noncommutative parameter as
\begin{equation}
G_{\a\b} = \e \eta_{\a\b},\ \  
\Theta^{01} = 2 \pi \tilde \a'_{\rm eff} \tilde G^2_{o(1)}.
\label{eq:od1p}
\end{equation}
As expected, we have $\tilde \a' \tilde g^{(1)}_s G^{\a\b} = \tilde \a'_{\rm
eff} \tilde G^2_{o(1)} \eta^{\a\b}$. 
This is a well-defined perturbative theory for small $\tilde G^2_{o(1)}$. 
The usual (1 + 3)-dimensional NCYM is believed to be renormalizable and
therefore it is a well-defined 
perturbative noncommutative field theory for small coupling 
$\tilde g^2_{\rm NCYM}$. Further we have $\tilde G^2_{o(1)} = 
\tilde g^2_{\rm NCYM}/(2\pi)$ which
implies that the two perturbative theories break down at the same time
when either of the couplings is strong. As mentioned earlier, the two
have basically the same gravity dual description. Note that the NCYM can have
T-duality, and therefore it is not really a field theory 
since it does not have a well-defined energy-momentum tensor. 
All these indicate that 
the usual (1 + 3)-dimensional NCYM and the (1 + 3)-dimensional
open D-string theory are just two different descriptions of the same
physics.

\noindent
{\bf Open D2 theory}: This theory is related to OM theory compactified 
on a small magnetic circle and provides a completion of the usual 
(1 + 4)-dimensional NCYM. We will discuss this case in detail in section 5.

\noindent
{\bf Open D3 theory}: The decoupling limit for this theory contains D5
branes in the presence of a near-critical 4-form worldvolume field
strength $H_{0123} = \frac{1}{(2 \pi)^3 \e \tilde \a'^2_{\rm eff}
\tilde G^2_{o(3)}} (1 - \frac{\e}{2})$. The bulk scalings are
\begin{eqnarray}
&&\tilde \a' = \e^{1/2} \tilde \a'_{\rm eff}, \ \
\tilde g^{(3)}_s = \tilde G^2_{o(3)},\ \ g_{\mu\nu} = 
\eta_{\mu\nu}\ (\mu, \nu = 0, 1, 2, 3),\nonumber\\
&&g_{ij} = \e \delta_{ij}, \ (i, j = 4, 5), \ \ g_{mn} = \e
\delta_{mn}~(m, n = {\rm transverse}), 
\label{eq:od3dl}
\end{eqnarray}
The coupling for this theory is related to the usual (1 + 5)-dimensional
NCYM coupling as
\begin{equation}
\tilde G^2_{o(3)} = \frac{\tilde g^2_{\rm NCYM} \tilde \a'^{- 1}_{\rm eff}}{(2\pi)^3}.
\end{equation}
The usual (1 + 5)-dimensional NCYM is nonrenormalizable and as such it
is  an effective theory. The present open D3 brane theory provides a
completion of this NCYM. Therefore this is an example that the open D3 brane
description is better than the usual NCYM one (or the F-string description). As
we will discuss this case further in section 6, this open D3 brane
theory is actually self-dual under S-duality.

In a similar fashion as discussed in \cite{gopmss}, different (1 +
 p)-dimensional open
D(p - 2) brane theories here can be related to each other either by a 
T-duality along a direction of the D(p - 2) brane or by a T-duality
along a direction transverse to both this D(p - 2) brane and the parent 
Dp-brane. However, a T duality along any codimension gives a D(p - 1)
brane which no longer lives inside the parent D(p - 1) brane. This
indicates that such a T-duality may render the open D(p - 1) brane
undecoupled. If we compactify the $x^{p - 2}$-direction with the
identification $x^{p - 2} \sim x^{p - 2} + 2 \pi R_{p - 2}$, the usual
transformations of bulk quantities under a T-duality along this direction  
give the following
\begin{equation}
H_{01\cdots (p - 3)} = 2 \pi R_{p - 2} H_{01\cdots (p - 2)}, \ \
 R'_{p - 2} = \frac{\tilde \a'_{\rm eff}}{R_{p - 2}}, \ \
\tilde G^2_{o(p - 3)} = \frac{\sqrt{\tilde \a'_{\rm eff}}}{R_{p -
2}}\tilde G^2_{o(p - 2)},
\label{eq:td}
\end{equation}
where $ R'_{p - 2}$ is the T-dual coordinate radius. One can check
that the resulting decoupling limit is for a (1 + (p -1))-dimensional
 open D(p - 3) brane theory.

\sect{(1 + p)-Dimensional Noncommutative Field Theories}

We follow the same steps as what we did in the previous section but now
for a (1 + p)-dimensional NCOS rather than for a (1 + p)-dimensional NCYM. 
From the open string perspective, the critical electric field limit
gives a (1 + p)-dimensional NCOS. The question is: what is the
corresponding decoupled theory with the same bulk scalings but now from
the open D(p - 2) brane perspective? As we will argue below, the answer
seems a decoupled (1 + p)-dimensional ``noncommutative'' field theory defined
on a noncommutative  geometry which is in general 
different from that for the usual (1 + p)-dimensional NCYM.

The decoupling limit for a (1 + p)-dimensional NCOS can be given
collectively as \cite{gopmss}:
\begin{eqnarray}
&&\a' = \e \a'_{\rm eff}, \ \ g_s = \frac{G^2_o}{\sqrt{\e}},\ \ 
g_{\mu\nu} = \eta_{\mu\nu}~(\mu, \nu = 0, 1), \nonumber\\
&& g_{ij} = \e
\delta_{ij}~(i, j = 2, \cdots p), \ \ g_{mn} = \e \delta_{mn}~ (m, n =
{\rm transverse}),\nonumber\\
&& 2\pi\a' \e^{01} F_{01} = 1 - \frac{\e}{2},
\label{eq:ncosdl}
\end{eqnarray}
where the scaling parameter $\e \to 0$ and the NCOS parameters $\a'_{\rm
eff}$ and $G_o$ remain fixed.

	The Dp brane worldvolume Poincare dual of $F_{01}$, i.e., 
$H_{2\cdots p}$, can be obtained, following the same steps as those 
given in the previous section, as
\begin{equation}  
\sqrt{- \det g} \frac{H_{2\cdots p}}{2\pi} = - \frac{1}{2}
\frac{\e_{2\cdots p \mu\nu}}{\sqrt{- \det g}} 
\frac{\partial {\cal L}_{DBI}}{\partial F_{\mu\nu}},
\label{eq:hdfone}
\end{equation}

Using the scaling limit given in (\ref{eq:ncosdl}) for the metric, the
closed string coupling and the near-critical electric field, we have
\begin{equation}
H_{2\cdots p} = \frac{1}{(2\pi)^{p - 2} \a'^{\frac{p - 1}{2}}_{\rm eff}
G^2_o},
\label{eq:mf}
\end{equation}
which remains fixed.

In summary, from the open D(p - 2) brane perspective, we have now
the following scaling limits:
\begin{eqnarray}
&&\a' = \e \a'_{\rm eff}, \ \ g_s = \frac{G^2_o}{\sqrt{\e}},\ \ 
g_{\mu\nu} = \eta_{\mu\nu}~(\mu, \nu = 0, 1), \nonumber\\
&& g_{ij} = \e
\delta_{ij}~(i, j = 2, \cdots p), \ \ g_{mn} = \e \delta_{mn}~ (m, n =
{\rm transverse}),\nonumber\\
&& H_{2\cdots p} = \frac{1}{(2\pi)^{p - 2} \a'^{\frac{p - 1}{2}}_{\rm eff}
G^2_o}.
\label{eq:ncfdl}
\end{eqnarray}

Let us inspect the action (\ref{eq:dp2a}) proposed in the previous
section  for
 the open D(p - 2) brane ending on the Dp brane which moves in the
 background given in (\ref{eq:ncfdl}). For convenience, we write it down
here as
\begin{equation}
S_{(p - 2)} = - \frac{1}{(2 \pi)^{p - 2} \a'^{(p - 1)/2} g_s}\int_{M^{p
- 1}} d^{p - 1} \sigma \sqrt{- \det \hat g_{\alpha\beta}} + 
\int_{M^{p - 1}} {\cal H}_{p - 1},
\label{eq:ndp2a}
\end{equation}   
where we have dropped the D(p - 2) brane worldvolume U(1) field for the
reason mentioned in the previous section, the D(p - 2) brane worldvolume
indices $\alpha, \beta = 0, 1, \cdots (p - 2)$ 
and the induced worldvolume metric 
\begin{equation}
\hat g_{\alpha\beta} = \partial_\alpha X^M \partial_\beta X^N g_{MN},
\end{equation}
where the metric $g_{MN}$ is the bulk spacetime one with 
$M, N = 0, 1, \cdots 9$. The above
Nambu-Goto-type action is not convenient for considering the scaling
behavior of the action. We here follow the procedure given
in~\cite{berbssone} to introduce the auxiliary worldvolume metric 
$\gamma_{\alpha\beta}$ and recast the above action in Polyakov form as
\begin{eqnarray}
S_{(p - 2)} &=& - \frac{1}{2 (2 \pi)^2 \a' g_s}\int_{M^{p
- 1}} d^{p - 1} \sigma \sqrt{- \det \gamma} \left(\gamma^{\alpha\beta}
\partial_\alpha X^M \partial_\beta X^N g_{MN} - 
(2 \pi)^2 (p - 3) \alpha'\right)\nonumber\\
&~& +  \int_{M^{p - 1}}  H_{p - 1},
\label{eq:rdp2a}
\end{eqnarray}
where we have again followed \cite{berbss} by insisting the worldvolume
coordinates $\sigma^\alpha$ as dimensionless. One can check that the
equation of motion for $\gamma_{\alpha\beta}$ gives the induced metric
and if substituting this back to the above action, we end up with the
Nambu-Goto action~(\ref{eq:ndp2a}). In the following, we consider the
scaling behavior of the above action under the scaling limit 
(\ref{eq:ncfdl}). As it is understood that the coordinates $X^M$ are now
fixed. The D(p - 2) brane coordinates $\sigma^\alpha$ as well as its
intrinsic metric $\gamma_{\alpha\beta}$ are also fixed. With these, we
have
\begin{eqnarray}
S_{(p - 2)} &&= - \frac{1}{2 (2\pi)^2 \a'_{\rm eff} G^2_o} \int_{M^{p -
1}} d^{p - 1} \sigma \sqrt{- \det \gamma} \left[\e^{- 1/2}\gamma^{\alpha\beta}
\partial_\alpha X^\mu \partial_\beta X^\nu \eta_{\mu\nu}
\right.\nonumber\\
&& \left. + \e^{1/2}      
\gamma^{\alpha\beta} \partial_\alpha X^i \partial_\beta X^j \delta_{ij}
+ \e^{1/2} \gamma^{\alpha\beta}
\partial_\alpha Y^m \partial_\beta Y^n \delta_{mn} - \e^{1/2}
(2\pi)^2 \a'_{\rm eff} (p - 3) \right]\nonumber\\
&& + \frac{1}{(p - 1)!}\int_{M^{p -
1}} d^{p - 1} \sigma \e^{\alpha_0 \alpha_1 \cdots \alpha_{p - 2}}
\partial_{\alpha_0} X^{i_1} \partial_{\alpha_1} X^{i_2} \cdots
\partial_{\alpha_{p - 2}} X^{i_{p - 1}} {\cal H}_{i_1 i_2 \cdots i_{p - 1}},
\label{eq:sb}
\end{eqnarray}
where $Y^m$ denote the bulk modes in directions 
transverse to the base Dp brane. From the above, we have the following:
\begin{enumerate}
\item The bulk modes $X^\mu$ for $\mu = 0, 1$ are frozen out.

\item The action for the bulk modes $X^i$ and $Y^m$  vanishes.
\end{enumerate}

Since the bulk field ${\cal H}_{2\cdots p}$ as given in~(\ref{eq:ncfdl})
is a fixed constant, the bulk theory is now described by the
following topological action
\begin{equation}
S_{(p - 2)} =   \frac{1}{(p - 1)!}\int_{M^{p -
1}} d^{p - 1} \sigma \e^{\alpha_0 \alpha_1 \cdots \alpha_{p - 2}}
\partial_{\alpha_0} X^{i_1} \partial_{\alpha_1} X^{i_2} \cdots
\partial_{\alpha_{p - 2}} X^{i_{p - 1}} 
{\cal H}_{i_1 i_2 \cdots i_{p - 1}},
\label{eq:wza}
\end{equation}
which in turn can be expressed as the following boundary action for $p
\ge 3$
\begin{equation}
\frac{1}{(p - 1)!} \int_{\partial M^{p - 1}} d^{p - 2} \xi  
\e^{\alpha_0 \alpha_1 \cdots \alpha_{p - 3}}
\partial_{\alpha_0} X^{i_1} \partial_{\alpha_1} X^{i_2} \cdots
\partial_{\alpha_{p - 3}} X^{i_{p - 2}}  X^{i_{p - 1}}{\cal H}_{i_1 i_2 
\cdots i_{p - 1}},
\label{eq:ba}
\end{equation}
where $\xi^\a$ with ($\a = 0, 1, \cdots (p - 3)$) denote now the local
coordinates for the boundary (p - 3)-brane and $X^i$ are the embedding
fields of the boundary (p - 3)-brane.

	The boundary degrees of freedom for the D(p - 2) brane are
governed by the above action. For $p = 2$, we can see that the action
(\ref{eq:wza}) has no local dynamics for a constant $H_i$. We therefore
don't expect the noncommutative geometry to arise for this case.
 For $p = 3$, 
the quantization of the above action gives $[X^i, X^j] \neq 0$,
therefore implying the spatial noncommutative geometry of the base D3-brane
along the line as for the usual NCYM discussed in \cite{seiw}.  For $p = 4, 5$,
we may follow~\cite{berbssone} to discuss the corresponding spatial
noncommutativity geometries of the base Dp-branes.  
However, for the $p = 5$ case, the S-dual of the resultant theory does
not appear to decouple from the bulk as we will discuss in section
6. This may indicate that the present theory is not well-defined,
either. For this reason, we postpone to study this case carefully
elsewhere, not pursuing it further in this paper. 
Therefore, except for the $p = 2$ case,
we expect in general that we have a noncommutative  
geometry for the base
Dp-brane upon the quantization of the above action. The remaining
question is: what is the decoupled theory at hand with the decoupling
limit (\ref{eq:ncfdl})?

	Our current knowledge is that a decoupled open brane theory 
requires usually a near-critical electric-like background field while a 
decoupled field theory requires a fixed magnetic-like background field 
(with respect to the fixed coordinates). With this, we might expect
that the decoupling limits (\ref{eq:ncfdl}) describe decoupled field
theories defined on noncommutative  geometries  
determined through the quantization of the action (\ref{eq:ba}).
Naively, we may take the field theory modes on Dp branes as super
Yang-Mills multiplet. This would imply that the above decoupled field
theories are also ``noncommutative'' Yang-Mills theories but now
defined on noncommutative geometries 
which are in general different from those for the usual NCYM. 

Given that the decoupled field theory is obtained from the open D(p - 2)
brane perspective and the noncommutative geometry is determined through
the fixed Dp-brane worldvolume $H_{p - 1}$-form, the resultant decoupled 
theory is naturally expected to be a tensor field theory since the field
theory modes on a single Dp brane is a tensor multiplet\footnote{For $p > 3$, 
 we know only how to deal with a single Dp brane since at present we
don't know how to generalize an abelian tensor multiplet to its
non-abelian one.} which is Poincare dual
to the U(1) gauge modes on the brane. If such a field theory for $p > 3$
exists indeed, the question is: Can we use the (1 + p)-dimensional
Poincare dual to map this decoupled field theory to a NCYM? 
To address this, we first need to know if it is
consistent to Poincare dual the dynamical tensor field while leaving the 
``noncommutative'' geometry intact. If this is true, 
we can end up with a
U(1) gauge field defined on a ``noncommutative'' geometry determined by the
boundary action (\ref{eq:ba}). If this is not true, we don't expect that
we can end up with a field theory since the Poincare dual of spatial
``noncommutative'' geometry would imply a time-space one. The expected
theory should be the (1 + p)-dimensional NCOS but we cannot get it by
performing the Poincare dual on the decoupled tensor field theory since
the later is expected to be an incomplete description of the underlying
physics while the former is a complete description for $p > 3$. 
Work on this issue for $p = 4$ case is in progress.

In spite of what has been said above, directly confirming the existence of 
the (1 + p)-dimensional ``noncommutative'' tensor field theories 
may not be easy since we need to know the effective open D(p - 2) brane metric
which is hardly available for $p > 3$. For $p = 3$, however, we are
reasonably sure that we end up with a (1 + 3)-dimensional noncommutative
Yang-Mills which is actually identical to  the usual (1 + 3)-dimensional
NCYM if their parameters are properly identified.

Let us give some detail about this theory. As discussed above, quantization of
the boundary action (\ref{eq:ba}) for $p = 3$  gives
\begin{equation}
[x^2, x^3] = - i 2\pi \a'_{\rm eff} G^2_o.
\end{equation}
Therefore, we have the spatial noncommutative parameter 
$\Theta^{23} = - 2 \pi \a'_{\rm eff} G^2_o$. The present decoupled theory
is obtained from the open D-string ending on D3-branes in the decoupling
limit (\ref{eq:ncfdl}) for $p = 3$. Given the relation between D-string
and F-string, we expect that the low energy Born-Infeld action for D3
branes with the open D-string ending on them can be obtained from that
for D3 branes with a F-string ending on them through the following 
replacements
\begin{equation}
g_s \to \frac{1}{g_s}, \ \ \a' \to \a' g_s, \ \ F_{\a\b} \to
H_{\a\b},
\end{equation}
where $F_{\a\b}$ is the worldvolume gauge field in the F-string
picture while $H_{\a\b}$ is the corresponding one in the D-string
picture. With the above, the decoupling limit (\ref{eq:ncfdl})
is essentially the same as the one for the usual NCYM as given 
in (\ref{eq:ncymdl}) in the previous section. Given the above,
let us make a consistent check on the
open D-string metric, the noncommutative parameter and the gauge
coupling using the
corresponding Seiberg-Witten relations for
the present noncommutative Yang-Mills theory. They are now
\begin{eqnarray}
&&G_{\a\b} = g_{\a\b} - (2\pi \a' g_s)^2 (H g^{-1} H)_{\a\b},\nonumber\\    
&&\Theta^{\a\b} = 2\pi \a' g_s \left(\frac{1}{g + 2\pi\a' g_s
H}\right)^{\a\b}_A,\nonumber\\
&& \frac{1}{g^2_{\rm NCYM}} = \frac{g_s}{2\pi} \left(\frac{\det (g + 2\pi
\a' g_s H)}{\det G}\right)^{1/2},
\end{eqnarray}
where $A$ in $()_A$ denotes the anti-symmetric part of the matrix.
Using the decoupling limit (\ref{eq:ncfdl}) for $p = 3$, we have from
the above
\begin{equation}
G_{\a\b} = \eta_{\a\b}, \ \ \Theta^{23} = - 2\pi\a'_{\rm eff} G^2_{o}, \
\ \ \frac{1}{g^2_{\rm NCYM}} = \frac{G^2_o}{2\pi}.
\label{eq:nncymp}
\end{equation} 
The noncommutative parameter $\Theta^{23}$ is the same as the one
obtained above and the open string metric is also expected. The
Yang-Mills coupling is inversely related to the open string coupling 
for NCOS. This is quite different from that between the open D-string
coupling and the usual NCYM coupling as given in (\ref{eq:odpc}) for 
$p = 3$. 

Under S-duality, we expect that our open D-string theory discussed in
 the previous section is mapped to the present NCOS via
\begin{equation}
\tilde \a'_{\rm eff} \to \a'_{\rm eff} = \tilde\a'_{\rm eff}
\tilde G^2_{o(1)},\ \ \tilde G^2_{o(1)} \to G^2_o = \frac{1}{\tilde 
G^2_{o(1)}},
\end{equation}
which are obtained from $\tilde \a' \to \a' = \tilde\a' \tilde g_s,\ \ 
\tilde g_s \to g_s = 1/{\tilde g_s}$.

With the above relation, we have the same parameters for the usual NCYM
and the above NCYM. Therefore, they are identical theories. In other
words, the NCYM keeps intact under S-duality. This is just the
consequence of S-duality given the two S-duality related bulk scalings
 and the relation $F_{23} = H_{23}$. In other words,  the low
energy dynamics of the open F-string ending on the base D3 branes with
background $F_{23}$ is identical to that of the open D-string
ending on the same D3 branes with background $H_{23}$.

	Note that the above S-duality for the NCYM is induced from that
for the bulk type IIB string theory. This is different from the usual
one which requires in addition a worldvolume Poincare 
duality for the background field. The usual S-duality maps the usual
NCYM directly to the NCOS as discussed in~\cite{gopmms}.
 In terms of our interpretation, the NCYM keeps 
intact under S-duality. 

At low energies, the NCOS, our open D-string theory and the NCYM are all
 expected to reduce to the corresponding usual Yang-Mills theories.
The question is: What are the relations among the three usual Yang-Mills 
theories. Let us find them out. For the NCOS, we have the gauge coupling
from~\cite{gopmss} as $g^2_{\rm YM} = 2\pi G^2_o$. For the NCYM, the
 gauge coupling is just $g^2_{NCYM} = \tilde g^2_{\rm NCYM} =
 2\pi G^2_{(1)} = 2\pi /G^2_o$. For our open D-string theory, 
we can calculate $g^2_{\rm YM} = 2\pi/G^2_{(1)}$. Given $G^2_o =
 1/G^2_{(1)}$, we have the same low energy Yang-Mills theory for the
 NCOS and our open D-string theory since the gauge coupling is the same.
However, we have $g^2_{\rm YM} = (2\pi)^2/g^2_{\rm NCYM}$. In other
 words, the low energy Yang-Mills theory from either the NCOS or our
 open D-string theory is strong-weakly related to that from the NCYM. 
This is the manifestation of the S-duality for the usual (1 + 3)-dimensional
 YM. This result is consistent with the S-duality relation between the
 NCOS and the usual NCYM discussed in~\cite{gopmms} even though our 
interpretation here is different as mentioned above.

\sect{Compactification of OM Theory on a Circle and
(1 + 4)-Dimensional Theories}

In this sub-section, we try to make connections of the (1 +
4)-dimensional open D2 brane theory and the new (1 + 4)-dimensional
NCYM discussed in the previous two sections to the compactification
of OM theory on a (either magnetic or electric) circle. We will see that 
the dimensional reduction of OM theory on either a magnetic circle or
an electric circle indicates the existence of the open D2 brane theory  
or the new (1 + 4)-dimensional NCYM. 

\subsection{OM Theory on a magnetic circle and 5-D Open D2 Brane Theory}  

In this section, we try to show that OM theory describes the strong
coupling of the usual (1 + 4)-dimensional open D2 brane theory discussed in
section 2. We also show that this open D2 brane theory provides a 
UV completion of the (1 + 4)-dimensional NCYM\footnote{That the UV completion
of the (1 + 4)-dimensional NCYM is an open D2 brane theory was also briefly
mentioned in a recent paper~\cite{ber}. An open D3 brane theory as the
UV completion of the (1 + 5)-dimensional NCYM was also mentioned
there. The author would like to thank R.-G. Cai for bringing his
attention to this reference.}.
 
As discussed in \cite{gopmss}, OM theory on a magnetic circle gives
NCYM with rank-2 noncommutative matrix with the following parameters
\begin{eqnarray}
&&\tilde \a' = \frac{1} {L \sqrt{2 M^2_{\rm eff} M^3_p}}, \ \ \tilde g_s =
\left(\frac{2 L^2 M^2_{\rm eff}}{M_p}\right)^{3/4}, \ \ \tilde g^2_{\rm NCYM} = 4
\pi^2 L, \nonumber\\
&&g_{\mu\nu} = \eta_{\mu\nu}~(\mu, \nu = 0, 1, 2), \ \ g_{ij} = 2 
\frac{M^3_{\rm eff}}{M^3_p} \delta_{ij},\ (i, j = 4, 5), \ \ F_{45} = 
\frac{L M^3_{\rm eff}}{\pi},
\label{eq:fivencymdl}
\end{eqnarray}
where $L$ is the coordinate radius of the magnetic circle, $M_{\rm eff}$
is the energy scale for the OM theory and $M_p$ is the
eleven-dimensional Planck scale which is sent to infinity in the
decoupling limit for OM theory. It was also concluded in that paper that
OM theory provides a completion of the (1 + 4)-dimensional NCYM. The
detailed path, as we show below, is that the (1 + 4)-dimensional open D2
brane found in this paper provides a completion of the NCYM and OM
theory describes the strong coupling of this open D2 brane theory.

Comparing the above with the decoupling
limit for NCYM with $p = 4$ in Eq. (\ref{eq:ncymdl}), we have also 
\begin{equation}
\e = 2 \frac{M^3_{\rm eff}}{M^3_p}, \ \ \tilde \a'_{\rm eff} = \frac{1}{2 L
M^3_{\rm eff}}
\label{eq:scalingp}
\end{equation}

	The (1 + 4)-dimensional NCYM is nonrenormalizable and therefore
this theory does not have a complete (1 + 4)-dimensional description.
 However,  when $L << 1/M_{\rm eff}$, the
magnetic circle is invisible to OM theory. We should end up with a 
(1 + 4)-dimensional open membrane theory which provides
a completion of the NCYM. We will show below that this open
membrane theory is our open D2 brane theory.

As discussed in the Introduction, an alternative description of this 
compactification of OM theory is via the open membrane since the 
compactification along the magnetic circle is transverse to the open 
membrane which is used in defining the OM theory. With this in
mind, we have from $\tilde g_s = \e^{1/4} \tilde G^2_{o(2)}$ and the relations given
in Eqs. (\ref{eq:fivencymdl}) and (\ref{eq:scalingp}) 
\begin{equation}
\tilde G^2_{o(2)} = \frac{\tilde g^2_{\rm NCYM} \tilde \a'^{- 1/2}_{\rm eff}}{(2\pi)^2} = 
\left(2 L M_{\rm eff}\right)^{3/2},
\label{eq:d2c}
\end{equation}
The scalings of other parameters for the OM theory can be
read from \cite{gopmss} as\footnote{Our convention here for $H_{012}$ 
differs from that used in~\cite{gopmss} by a factor of $(2\pi)^2$.}
\begin{eqnarray}
&&\tilde \a' = \e^{1/2} \tilde \a'_{\rm eff}, \ \ \e = e^{- 2\b} = 
2 \frac{M^3_{\rm eff}}{M^3_p}\nonumber\\
&& g_{\mu\nu} = \eta_{\mu\nu}~ (\mu, \nu
= 0, 1, 2), \ \ g_{ij} = \e \delta_{ij}~ (i, j = 4, 5),\nonumber\\
&& H_{012} = \frac{M^3_p \tanh\b}{(2 \pi)^2} = \frac{1}{(2\pi)^2
\tilde \a'^{3/2}_{\rm eff} \tilde G^2_{o(2)}} \left(\frac{1}{\e} - \frac{1}{2}\right).
\end{eqnarray}
The above parameters and scalings are precisely what we used to
define our (1 + 4)-dimensional open D2 brane theory in section 3. 
If we examine the
coupling $\tilde G_{o(2)}$ of our open D2 brane theory given (\ref{eq:d2c}), we
have $\tilde G_{o(2)} << 1$ if $ L << 1/M_{\rm eff}$ and $\tilde G_{o(2)} >> 1$ if $L
>> 1/M_{\rm eff}$. The former implies that the magnetic circle is
invisible to OM theory while the latter says that the circle appears to be
uncompactified to OM theory.
Therefore, our
open D2 brane theory is OM theory on a magnetic circle when $L <<
1/M_{\rm eff}$ and provides a completion of the usual (1 + 4)-dimensional
NCYM. Its strong coupling is OM theory.

	In summary, when $L << 1/M_{\rm eff}$ and the relevant energy
scale $ << 1/\tilde g^2_{\rm NCYM}$, both OM theory and our open D2 brane theory
can be effectively described by the usual (1 + 4)-dimensional NCYM. When 
we have only $L << 1/M_{\rm eff}$, OM theory reduces to our open D2
brane theory. In other words, OM theory provides a  completion
of our open D2 brane in coupling while our open D2 brane provides an
completion of the usual (1 + 4)-dimensional NCYM in energy.

\subsection{OM theory on an electric circle and 5-D noncommutative
tensor field theory}

As discussed in~\cite{gopmss}, the compactification of OM theory on an
electric circle (say in the 2 direction) with proper (also coordinate)
radius $R$ gives (1 + 4)-dimensional NCOS with the following
parameters:
\begin{eqnarray}
&&\a' = \frac{1}{R M^3_p}, \ \ g_s = (R M_p)^{3/2}, \ \ 2 \pi \a' F_{01}
= \a' R H_{012} = 1 - \frac{M^3_{\rm eff}}{M^3_p},\nonumber\\
&& g_{\mu\nu} = \eta_{\mu\nu}~(\mu, \nu = 0, 1), \ \ g_{ij} = \frac{2
M^3_{\rm eff}}{M^3_p} \delta_{ij}~(i, j = 3, 4, 5),\nonumber\\
&& g_{mn} = \frac{2
M^3_{\rm eff}}{M^3_p} \delta_{mn}, ~(m, n = {\rm transverse}),
\label{eq:5dncosdl}
\end{eqnarray}
where $M_p \to \infty$ is understood. Comparing with the  decoupling
limit for NCOS given in~(\ref{eq:ncosdl}) for $p = 4$, we have
\begin{equation}
\e = \frac{2 M^3_{\rm eff}}{M^3_p}, \ \ \a'_{\rm eff} = \frac{1}{2 R
M^3_{\rm eff}}, \ \ G^2_o = \sqrt{2} (R M_{\rm eff})^{3/2}.
\label{eq:ncosp}
\end{equation}

It is not difficult to see that $G_o >> 1$ implies $R >> 1/M_{\rm eff}$.
In other words, the circle appears uncompactified. Therefore, OM theory
provides a completion of the (1 + 4)-dimensional NCOS in coupling. On
the other hand, if $G_o << 1$, we have $R << 1/M_{\rm eff}$. This is to say
that the circle is invisible to OM theory. Since one of the dimensions
of the open membrane in OM theory is wrapped on this circle, we therefore
end up with the above NCOS theory. 

	Again as discussed in the Introduction, we can instead focus on
the magnetic 3-form field $H_{345}$ rather than on the electric one. The 
question is: what is the decoupled theory in this case? Let us examine
the decoupling limit. Since the change here is to replace $F_{01}$
by $H_{345}$, we therefore have the following:
\begin{eqnarray}
&&\a' = \e \a'_{\rm eff}, \ \ g_s = \frac{G^2_o}{\sqrt{\e}}, \ \ H_{345} =
 \left(\frac{2 M^3_{\rm eff}}{M_p}\right)^{3/2}
\frac{\sinh\beta}{(2\pi)^2} =  \frac{ 2 M^3_{\rm eff}}{(2\pi)^2} = 
 \frac{1}{(2\pi)^2 \a'^{3/2}_{\rm eff} G^2_o}, 
\nonumber\\
&& g_{\mu\nu} = \eta_{\mu\nu}~(\mu, \nu = 0, 1), \ \ g_{ij} = 
\e \delta_{ij}~(i, j = 3, 4, 5),\nonumber\\
&& g_{mn} = \e \delta_{mn}, 
~(m, n = {\rm transverse}),
\label{eq:5dncfdl}
\end{eqnarray}
where the parameters $\e, \a'_{\rm eff}$ and $G_o$ are given 
in~(\ref{eq:ncosp}). Note that our convention for the above $H_{345}$
 differs from that given in \cite{gopmss}: our $H_{234}$ corresponds to 
$ - H_{345} / (2 \pi)^2$ used in~\cite{gopmss}. 
With this in mind, the above limit gives
precisely the one in (\ref{eq:ncfdl}) for $p = 4$. As discussed in 
the previous section, this limit gives a (1 + 4)-dimensional
 tensor field theory defined on a
noncommutative  geometry which is determined upon 
the quantization of the boundary action (\ref{eq:ba}). 

	This (1 + 4)-dimensional tensor field theory is expected
to be  an effective theory and its completion is the (1 + 4)-dimensional 
NCOS.

\sect{Relation to ODp Theories from NS5-branes}

	As discussed in the Introduction, the existence of ODp theories
from NS5 branes for $p \le 5$, as discovered independently in 
\cite{gopmss,hartwo}, can be traced back to the fact that an open 
Dp brane can end on NS5 branes. These ODp theories are also related to
the known NCOS theories (for example, the (1 + 5) NCOS) and to each other
through S- and T-dualities \cite{gopmss}.

	The scaling limits for these ODp are given in \cite{gopmss} as
\begin{eqnarray}
&&  \bar\a' = \e^{1/2} \bar \a'_{\rm eff}, \ \ g_s^{(p)} = \e^{(3 - p)/4}
 \bar G^2_{o(p)}, \ \ g_{\mu\nu} = \eta_{\mu\nu},~(\mu, \nu  = 0, 1, \cdots,
 p), \nonumber\\
&& g_{ij}
 = \e \delta_{ij}~(i, j = (p + 1), \cdots 5),\ \ g_{mn}
= \e \delta_{mn}, ~ (m, n = {\rm transverse}),\nonumber\\
&&\e^{01\cdots p} C_{01\cdots p} = \frac{1}{(2 \pi)^p \bar G^2_{o(p)} 
\bar \a'^{(p + 1)/2}_{\rm eff}} \left(\frac{1}{\e} -
\frac{1}{2}\right),\nonumber\\
&&C_{(p + 1) \cdots 5} = \frac{1}{(2 \pi)^{4 - p} \bar G^2_{o(p)}  \bar 
\a'^{(5 - p)/2}_{\rm eff}}.
\label{eq:nsodpdl}
\end{eqnarray}

In the above, both a RR (1 + p)-form and a RR (5 - p)-form potentials
are included for defining the ODp. These constant RR potentials can be
traded to the corresponding NS5 brane worldvolume (1 + p)-form 
field strength $H_{01 \cdots p}$ and (5 - p)-form field strength
$H'_{(5 - p) \cdots 5}$. Given the fact that the two are related to each
other by the worldvolume Poincare duality for $p = 2$ case, we expect
that the two are related so for a general $p \le 5$. In other words, the
(1 + p)-form field strength $H_{1 + p}$ and the (5 - p)-form field
strength $H'_{(5 - p)}$ are not independent to each other but related by
the worldvolume Poincare duality. This is consistent with the low energy
field contents on a NS5 brane in either IIA or IIB string theory for
which we don't have such two independent field strengths living on the
NS5 brane worldvolume at the same time. 
To avoid doubly counting degrees of freedom, we 
allow only one of them
present at one time except for the case of $p = 2, 5$. For the $p = 2$ case,
we still have only one 3-form field strength but with two nonvanishing
components related to each other by the non-linear worldvolume Poincare
duality. For the $p = 5$ case, neither the 6-form field strength nor the
the 0-form one carries local dynamics on the NS5 brane. For this reason,
they are allowed to present at the same time.
We therefore interpret that the decoupling limit for ODp given 
in~\cite{gopmss} should include only the $C_{01\cdots p}$ not the $C_{(p
+ 1)\cdots 5}$ one except for $p = 2, 5$ cases. This will affect the
interpretations for some of the ODp theories given in~\cite{gopmss}.

The properties for each of the ODp theories have been discussed
in~\cite{gopmss}. However, for $p = 3$, we interpret the OD3 theory
to be self-dual rather than to be S-dual to the usual (1 +
5)-dimensional NCYM as given in~\cite{gopmss}. In addition to
 the reason mentioned above, the other favoring our interpretation is
that the OD3 theory is a complete description while the (1 +
5)-dimensional NCYM is merely an effective one.
 We cannot expect that a complete theory is
mapped to an incomplete one under S-duality. This case is quite
different from that in  (1 + 3)-dimensions where the NCYM is also a
complete theory.  

 For different ODp, the origin of
the worldvolume background field $H_{01\cdots p}$ is different. Let us
explain this briefly. For $p = 0$, the D0 brane used in defining OD0
theory couples to a 1-form
field strength. This 1-form must be a derivative of one of the five 
scalars in the (2, 0) tensor multiplet.  Since this scalar interacts
with D0 brane charge and therefore must be the
zero mode associated with the compactified direction transverse to the
original M5 brane which is now the NS5 brane in IIA. The Poincare dual
of this 1-form field strength on the NS5 brane worldvolume gives a
5-form field strength whose potential couples to the boundary of the
open D4 brane ending on the NS5 brane. The critical electric field limit
of this 5-form field strength, which is actually Poincare dual to a
magnetic-like 1-form $H_5$,  defines the OD4 theory. For even $p$, only 
the OD2 theory is defined as the critical field limit of the self-dual
field strength $H_{012}$ in the (2, 0) tensor multiplet. 

	For odd $p$, the NS5 brane is in Type IIB string theory. The low
energy field content on the NS5 brane is the (1, 1) vector multiplet.
The OD1 theory results from the critical
electric field strength $H_{01}$ whose potential is in the (1, 1) vector
multiplet. The OD3 theory results from a near-critical 4-form field
strength $H_{0123}$ which is Poincare dual to the magnetic-like 2-form
field strength $H_{45}$. So the origin of this 4-form field strength is
also clear. However, we have neither a 6-form field strength nor a
0-form field strength in the (1, 1) vector multiplet. Actually, a 6-form
or a 0-form field strength in (1 + 5)-dimensions carries no local
dynamics. For this reason, both of the 6-from and the 0-form can appear
at the same time. So for OD5, we can also have both the 6-form 
$H_{012345}$ and a 0-form $H$. Because of this, we don't have a
well-defined S-dual of OD5 as discussed in~\cite{gopmss}.

	One of purposes in this section is to show that the open Dp
brane and the NCYM theories discussed in section 3 and 4 are also
implied by the ODp theories given our above interpretation for the
NS5-brane worldvolume fields. For convenience, we rewrite the scaling
limits for ODp except for $p = 2, 5$ case using our interpretation as
\begin{eqnarray}
&&  \bar\a' = \e^{1/2}  \bar\a'_{\rm eff}, \ \ g_s^{(p)} = \e^{(3 - p)/4}
 \bar G^2_{o(p)}, \ \ g_{\mu\nu} = \eta_{\mu\nu},~(\mu, \nu  = 0, 1, \cdots,
 p), \nonumber\\
&& g_{ij}
= \e \delta_{ij}~(i, j = (p + 1), \cdots, 5), \ \ g_{mn} = 
\e \delta_{mn} ~ (m, n = {\rm transverse}),\nonumber\\
&&\e^{01\cdots p} H_{01\cdots p} = \frac{1}{(2 \pi)^p \bar G^2_{o(p)} 
 \bar \a'^{(p + 1)/2}_{\rm eff}} \left(\frac{1}{\e} -
\frac{1}{2}\right),
\label{eq:nnsodpdl}
\end{eqnarray}

	Let us point out first that except for the dimensionality 
(here it is (1 + 5)-dimensions), the scalings for the OD(p - 2) theories
in Eq. (\ref{eq:nnsodpdl}) look exactly the same as those for our 
(1 + p)-dimensional open D(p - 2) brane theories  discussed in
section 3 for $p \le 5$. We now explore the connection between these two.

	For this purpose, let us consider $p = 3$ in
Eq. (\ref{eq:nnsodpdl}). The decoupling limit for this OD3 is
\begin{eqnarray}
&& \bar \a' = \e^{1/2} \bar \a'_{\rm eff}, \ \ g_s^{(3)} = \bar G^2_{o(3)}, \ \
g_{\mu\nu} = \eta_{\mu\nu},~(\mu, \nu  = 0, 1, \cdots, 3), \nonumber\\
&& g_{ij}
= \e \delta_{ij}~(i, j = 4, 5),\ \ g_{mn}
= \e \delta_{mn}, ~ (m, n = {\rm transverse}),\nonumber\\
&&\e^{0123} H_{0123} = \frac{1}{(2 \pi)^2  \bar G^2_{o(3)}
 \bar \a'^{2}_{\rm eff}} \left(\frac{1}{\e} - \frac{1}{2}\right).
\label{eq:nsod3dl}
\end{eqnarray}

	If we S-dual this OD3 theory, we end up with another OD3
theory whose scalings look identical to the original ones except
for some changes for the fixed parameters $\bar \a'_{\rm eff},
 \bar G^2_{o(3)}$. This is due to the fact that the D3 brane is intact under 
S-duality\footnote{This is manifest by the fact that the near-critical 
electric field $H_{0123}$ is intact under S-duality. This becomes more
clear if we use $C_{0123}$ rather than the worldvolume $H_{0123}$.}. 
The only possible 
effects associated with the base NS5 brane in the decoupling limit are
on the closed 
string constant $ \bar\a'$ and the closed string coupling $g^{(3)}_s$. 
It turns out
that their scalings remain the same under S-duality for this case, a
welcome and yet expected result. If we denote
with $\tilde A$  as the  S-dual of quantity $A$ which is not invariant
under S-duality, we have
\begin{equation}
\bar\a' \to \tilde\a' =  \bar \a' g^{(3)}_s = \e^{1/2} \tilde\a'_{\rm eff}, \ \ 
g^{(3)} \to  \tilde g^{(3)} = \frac{1}{g^{(3)}} = \frac{1}{\bar
G^2_{o(3)}} = \tilde G^2_{o(3)},
\label{eq:sdualp}
\end{equation}
for which we insist that the closed string metric remains the same 
as before\footnote{
The notion that the string constant $ \a'$ transforms under S-duality
is due to our choice that the asymptotic string-frame metric does not
change under S-duality. This is an effective way in implementing
S-duality which is also useful. The original S-duality requires 
the Einstein-frame metric and $ \a'$ to be invariant under S-duality. 
 Let us demonstrate the above two cases in the following
simple examples: a) If we insist that the asymptotic string metric
remain the same but the $\a' \to \tilde\a' = \a' g_s$, we have
$(1/\a')\int \partial X^M \partial X^N g_{MN} \to (1/\tilde \a') \int
 \partial X^M \partial X^N g_{MN}$ under S-duality. This basically
says that a fundamental string with its parameter $\a'$ is mapped to
another fundamental string with its parameter $\tilde\a' = \a' g_s$.
However, if we interpret this new string in its original $\a'$, it is 
a D-string.
b) If we insist that only Einstein metric and $\a'$ remain invariant
under S-duality, we have $(1/\a') \int  \partial X^M \partial X^N
e^{\phi/2} g^E_{MN} \to (1/\a') \int  \partial X^M \partial X^N               
e^{- \phi/2} g^E_{MN} = (1/ g_s \a') \int  \partial X^M \partial X^N               
 g_{MN}$ where we have used the relation $g = e^{\phi/2} g^E$  
in relating the original string-frame metric $g$ to its Einstein-frame metric
$g^E$ in the last step. We have also used $\phi \to - \phi$ under
S-duality. If we interpret this string in the original string metric,
this S-dual string is a D-string because of the tension is now $\sim
1/(\a' g_s)$. However, it is still a fundamental string if we use the
S-dual string metric which is now $\tilde g = g/g_s$. Therefore the
above two pictures don't lead to any inconsistency. It is merely a choice
of attributing the change to the metric or to the string constant
$\a'$.}. This also implies that the D3 brane tension 
$\sim 1/ (\bar\a'^2 g^{(3)}_s)$ remains invariant under S-duality, again a
welcome and yet expected result. This further implies that the OD3
tension $\sim 1/(\bar\a'^2_{\rm eff} \bar G^2_{o(3)})$ also remains invariant
under S-duality which is consistent with the fact that $H_{0123}$ 
(or $C_{0123}$) is
intact under S-duality. Given that the closed string metric, 
 the proper tension of the D3-brane ending on the 
NS5 brane and  
 the near-critical electric field $C_{0123}$ all remain unchanged under
S-duality, we therefore still have an open D3 brane theory under
S-duality as claimed above with the following decoupling limit:
\begin{eqnarray}
&& \tilde \a' = \e^{1/2} \tilde\a'_{\rm eff}, \ \ \tilde g^{(3)}_s =
\tilde G^2_{o(3)}, \ \ g_{\mu\nu} = \eta_{\mu\nu} ~ (\mu, \nu = 0, 1, 2, 3), 
\nonumber\\
&& g_{ij} = \e 
\delta_{ij}~(i, j = 4, 5),\ \ g_{mn} = \e 
\delta_{mn}~ (m, n = {\rm transverse}),\nonumber\\
&&\e^{0123} H_{0123} = \frac{1}{(2 \pi)^2 \tilde G^2_{o(3)}
\tilde\a'^{2}_{\rm eff}} \left(\frac{1}{\e} - \frac{1}{2}\right),
\label{eq:od3dl}
\end{eqnarray}
where we have
\begin{equation}
\tilde\a'_{\rm eff} = \bar \a'_{\rm eff} \bar G^2_{o(3)}, \ \ \tilde
G^2_{o(3)} = 
\frac{1}{\bar G^2_{o(3)}},
\label{eq:od3pr}
\end{equation}
which implies $\tilde\a'^2_{\rm eff} \tilde G^2_{o(3)} = \bar \a'^2_{\rm eff}
 \bar G^2_{o(3)}$.
This new open D3 brane theory has the same tension as the original
 one but its coupling $\tilde G^2_{o(3)}$ is inversely related to the
 original one as indicated above. Therefore when one OD3 theory is
 strongly coupled, the other is weakly coupled and vice-versa. This new
 OD3 theory is just the open D3 brane theory discussed in section
 3. Subsequent applications of T-duality on this OD3 theory along
 $x^3, x^2, x^1$ as described in section 3 will give our open Dp brane
theories for $p \le 3$. Therefore, the OD(p - 2) 
theories from NS5 branes also imply the existence of those (1 + 
 p)-dimensional open D(p - 2) brane theories discovered in this paper.  

It is clear now that the OD(p - 2) theories from NS5 branes and those found in
this paper are U-duality related. Let us  make some further comparisons 
between them. First for $p \le 5$, our open D(p - 2) brane
theories live in (1 + p)-dimensions while those from NS5 brane always
live in (1 + 5)-dimensions. Assuming the respective compactification
radii to be the same, we have the ratio $\tilde G^2_{o(p - 2)}/\bar G^2_{o(p
- 2)} = 
1/\bar G^{p - 1}_{o(3)}$. If $\bar G_{o(3)} > 1$, then $\tilde G_{o(p)} <
\bar G_{o(p)}$ and the other way around if $\bar G_{o(3)} < 1$. Further 
$\tilde G^2_{o(p - 2)} \tilde \a'^{(p - 1)/2}_{\rm eff} = \bar G^2_{o(p - 2)} 
\bar \a'^{(p - 1)/2}_{\rm eff}$. This implies that our open D(p-2) brane
theory and that from NS5 brane have the same proper tension and the same
near-critical electric field $H_{01\cdots (p - 2)}$. The bulk 
 metric in both cases
remain the same. Therefore, the reason that our open D(p - 2) brane
theory can only see (1 + p)-dimensions while those from NS5 brane always
see (1 + 5)-dimensions may be due to the difference in their couplings.

For $ p = 5$, as discussed in section 3, the open D3
brane brane provides a completion of the usual (1 + 5)-dimensional NCYM.
Our discussion above says that the S-duality of this open D3 brane
theory is the OD3. This indicates that the S-duality of the usual
(1 + 5)-dimensional NCYM gives another (1 + 5)-dimensional NCYM.
 It is for this case that our
interpretation differs from that given in~\cite{gopmss} where the
S-duality of OD3 was interpreted to give the usual (1 + 5)-dimensional
NCYM. The question is: what is the new (1 + 5)-dimensional NCYM? This is
the topic to which we turn next.

Following the discussion given in section 3 and 4, we expect that we
 might have noncommutative field theories for $p = 0, 1, 3,
4$ if the open D(4 - p) brane description is insisted with the following
scaling limits
\begin{eqnarray}
&& \bar \a' = \e^{1/2} \bar \a'_{\rm eff}, \ \ g_s^{(p)} = \e^{(3 - p)/4}
\bar G^2_{o(p)}, \ \ g_{\mu\nu} = \eta_{\mu\nu},~(\mu, \nu  = 0, 1, \cdots,
p), \nonumber\\
&& g_{ij}
= \e \delta_{ij}~(i, j = (p + 1), \cdots 5), \ \ g_{mn}
= \e \delta_{mn}, ~ (m, n = {\rm transverse}),\nonumber\\
&&H_{(p + 1) \cdots 5} = \frac{1}{(2 \pi)^{4 - p}\bar G^2_{o(p)}\bar \a'^{(5 -
p)/2}_{\rm eff}}.
\label{eq:nsncfdl}
\end{eqnarray}

Let us examine the action of open D(4 - p) brane ending on NS5 branes: 
\begin{eqnarray}
S_{(4 - p)} &=& - \frac{1}{2 (2 \pi)^2 \bar\a' g^{(p)}_s}\int_{M^{5 - p}} d^{5 -
p} \sigma \sqrt{- \det \gamma} \left(\gamma^{\alpha\beta}
\partial_\alpha X^M \partial_\beta X^N g_{MN} - 
(2 \pi)^2 (3 - p) \alpha'\right)\nonumber\\
&~& +  \int_{M^{5 - p}}  H_{5 - p}.
\label{eq:nsfa}
\end{eqnarray}

With the scaling limits (\ref{eq:nsncfdl}), we have
\begin{eqnarray}
S_{(4 - p)} &=& - \frac{1}{2 (2 \pi)^2 \bar\a'_{\rm eff} \bar G^2_{o(p)}}
\int_{M^{5 - p}} d^{5 - p} \sigma \sqrt{- \det \gamma} \left[ \e^{- (p -
5)/4} \gamma^{\alpha\beta}\partial_\alpha X^\mu \partial_\beta X^\nu 
\eta_{\mu\nu}\right. \nonumber\\
&~&\e^{(p - 1)/4} \gamma^{\alpha\beta}\partial_\alpha X^i \partial_\beta
X^j \delta_{ij} +  \e^{(p - 1)/4} 
\gamma^{\alpha\beta}\partial_\alpha Y^m \partial_\beta Y^n \delta_{mn}\nonumber\\
&~&\left.- \e^{(p - 3)/4} (2 \pi)^2 (3 - p) \right]
+  \int_{M^{5 - p}}  H_{5 - p}.
\label{eq:nsfs}
\end{eqnarray}
where we denote $Y^m$ as the bulk modes along the directions transverse
to the NS5 brane.

Except for the $p = 1$ case, the only finite part of the above action is the 
bulk topological term which can be expressed in terms of the following
boundary action (except for the $p = 4$ case)
\begin{equation}
\frac{1}{(5 - p)!} \int_{\partial M^{5 - p}} d^{4 - p} \xi  
\e^{\alpha_0 \alpha_1 \cdots \alpha_{3 - p}}
\partial_{\alpha_0} X^{i_1} \partial_{\alpha_1} X^{i_2} \cdots
\partial_{\alpha_{3 - p}} X^{i_{4 - p}}  X^{i_{5 - p}}{\cal H}_{i_1 i_2 
\cdots i_{5 - p}}.
\label{eq:nsba}
\end{equation}

In other words, we can have noncommutative field theories for $p = 0,
3$ upon the quantization of the above action which determines the
geometry of the base NS5 brane. For $p = 0$, this appears to be a
noncommutative (2, 0) theory. Since the background field used in
defining this theory comes from the magnetic dual of the derivative of
the scalar in (2, 0) theory, whether we indeed have such a
noncommutative field theory needs further investigation. For $p = 3$, we
end up with the aforementioned NCYM which can actually be identified with 
the usual NCYM. We will show this later on. 

The $p = 4$ case does not give noncommutativity and therefore we expect
that we end up with the usual (2, 0) theory. For $p = 1$, the bulk modes
$X^i, Y^m$ remain even with the decoupling limit. This may indicate that
we don't have a decoupled noncommutative field theory. 
This also indicates that the (1 + 5)-dimensional noncommutative tensor
field theory discussed in
section 4 may not be well-defined either since it is expected to be
related to the present one by S-duality.

	We now discuss the $p = 3$ case mentioned above. The
quantization of the boundary action (\ref{eq:nsba}) for this case gives
\begin{equation}
[x^4, x^5] = - i 2\pi \bar\a'_{\rm eff} \bar G^2_{(3)},
\end{equation}
which gives the noncommutative $\Theta^{45} = - 2\pi \bar\a'_{\rm eff}
\bar G^2_{(3)}$.

	Given the S-dual relation between the open F-string ending on D5
branes and open D-string ending on NS5 branes, we expect, as before,
that the open D-string metric, the noncommutative parameter and the
gauge coupling can be calculated with the scaling limit (\ref{eq:nsncfdl})
  using the following Seiberg-Witten relations:
\begin{eqnarray}
&&G_{\a\b} = g_{\a\b} - (2\pi \bar \a' g^{(3)}_s)^2 (H g^{-1} H)_{\a\b},\nonumber\\    
&&\Theta^{\a\b} = 2\pi \bar\a' g^{(3)}_s \left(\frac{1}{g + 2\pi\bar\a' g^{(3)}_s
H}\right)^{\a\b}_A,\nonumber\\
&& \frac{1}{g^2_{\rm NCYM}} = \frac{g^{(3)}_s}{(2\pi)^3 (\bar\a'
g^{(3)}_s) } \left(\frac{\det (g + 2\pi
\bar\a' g^{(3)}_s H)}{\det G}\right)^{1/2},
\end{eqnarray}
where $\a,\b = 0, 1, \cdots, 5$. 
We find 
\begin{equation}
G_{\a\b} = \eta_{\a\b}, \ \ \Theta^{45} = - 2\pi \bar\a'_{\rm eff}
\bar G^2_{(3)},\ \  g^2_{\rm NCYM} = (2\pi)^3 \bar\a'_{\rm eff}.
\end{equation}
The fixed open D-string metric indicates that we indeed end up with
a noncommutative field theory. The noncommutative parameter is the same
as the one calculated above from the quantization of the boundary
action. Let us understand the above Yang-Mills coupling. Since an open
D-string ending on NS5 branes are S-dual to an open F-string ending on
D5 branes, we expect that the bulk scaling limits for this NCYM are
S-dual to those for the usual NCYM. This further implies that the
parameters for the two decoupled NCYM are related to each other. Let us
find these relations. The scaling limits for the usual (1 +
5)-dimensional NCYM  are given in
(\ref{eq:ncymdl}). Under S-duality, we have
\begin{equation}
\tilde \a' \to \bar\a' = \tilde \a' \tilde g^{(3)}_s, \ \ \tilde
g^{(3)}_s \to g^{(3)}_s =
\frac{1}{\tilde g^{(3)}_s}.
\end{equation}
From the above, we have
\begin{equation}
\bar G^2_{o(3)} = \frac{(2\pi)^3  \tilde \a'_{\rm eff}}{
\tilde g^2_{\rm NCYM}} = \frac{1}{\tilde G^2_{(3)}},\ \ \bar\a'_{\rm eff} = 
\frac{\tilde g^2_{\rm NCYM}}{(2\pi)^3} = \tilde \a'_{\rm eff} \tilde G^2_{(3)}.
\end{equation}
With this, we have 
\begin{equation}
\Theta^{45} = 2\pi \bar \a'_{\rm eff} \bar G^2_{o(3)} = 2\pi \tilde \a'_{\rm eff},
\ \ g^2_{\rm NCYM} = (2\pi)^3 \bar \a'_{\rm eff} = \tilde g^2_{\rm NCYM}.
\end{equation}
In other words, the two NCYM theories have the same parameters and 
they can actually be identified. Again this is just the consequence of 
S-duality. We have seen this for the two (1 +
3)-dimensional NCYM theories discussed in section 4. In other words, the NCYM
keeps intact under S-duality. 

At low energies, all these (1 + 5)-dimensional decoupled theories (i.e., 
the NCOS, OD1, OD3, our open D3 brane theory and the NCYM) from
type IIB string theory are expected to give the usual (1 +
5)-dimensional Yang-Mills. The question is: Can we have a unique usual
Yang-Mills? We can check this at least for the NCOS, OD1 and the NCYM. 
For the NCYM, from the above, we can see that the low energy limit can
be achieved by insisting $\tilde \a'_{\rm eff} \to 0$ while keeping
$\bar \a'_{\rm eff}$ fixed. This in turn implies that we set $\bar
G^2_{(3)} \to 0$.

For the NCOS, it reduces to the usual Yang-Mills with gauge coupling
$g^2_{\rm YM} = (2\pi)^3 G^2_o \a'_{\rm eff}$ as given in~\cite{gopmss}.
For the OD1, it reduces to
\begin{eqnarray}
S &=& \frac{\bar G^2_{(1)}}{4 (2\pi)^3 \bar \a' g^{(1)}_s} \int d^6 x
\sqrt{- G} G^{AC} G^{BD} \hat H_{AB} \hat H_{CD},\nonumber\\
&= & \frac{1}{4 (2\pi)^3 \bar\a'_{\rm eff}} \int d^6 x \eta^{AC}
\eta^{BD} \hat H_{AB} \hat H_{CD},
\end{eqnarray}
where the open D-string metric $G_{AB} = \e \eta_{AB}$ has been
used. From the above, we have $g^2_{\rm YM} = (2\pi)^3 \bar \a'_{\rm
eff}$. 

Since the NCOS (with parameters $\a'_{\rm eff}, G_o$) is S-dual to OD1
(with parameters $\bar\a'_{\rm eff}, \bar G_{(1)}$), we have the
following
\begin{equation}
\bar G^2_{(1)} = \frac{1}{G^2_o},\ \ \bar\a'_{\rm eff} = \a'_{\rm eff}
G^2_o.
\end{equation}
This implies that the low energy Yang-Mills theories from the above
three different theories are actual the same since the gauge coupling is
the same. This is different from the (1 + 3)-dimensional case 
discussed at the end of section 4.

\sect{(1 + 3)-Dimensional Open (p, q)-String Theory}

The discussion given in the previous sections hints already that 
we have interesting story in (1 + 3)-dimensions. For example, 
our (1 + 3)-dimensional open D-string theory discussed in section
3 is equivalent to  the usual (1 + 3)-dimensional NCYM. 
 We intend to give explanations
for related issues in this section.  

In \cite{gopmms}, it was shown that the S-duality of (1 + 3)-dimensional 
NCYM gives (1 + 3)-dimensional NCOS. This conclusion, in spite of its 
correctness, does raise the following puzzles: a) Why is this true 
only for the (1 + 3)dimensional NCYM, not for the
(1 + 5)-dimensional one, for example? b) How can we reconcile this with 
the belief that the non-perturbative quantum SL(2, Z) symmetry
of the parent type IIB string theory is actually inherited to its
decoupled sub-theory (we call it the little type IIb string theory)
without gravity? 

As we know that the existence of D-string or in general
a (p, q)-string is a consequence of this SL(2, Z) symmetry in the 
non-perturbative type IIB string theory. By the same token, if we have 
SL(2,Z) symmetry for the little type IIb string theory, the existence of
(1 + 3)-dimensional NCOS should imply a (1 + 3)-dimensional open D-string
or in general a (1 + 3)-dimensional open (p, q)-string theory. However,
the above conclusion given in~\cite{gopmms} says that the S-dual of the
NCOS is the usual (1 + 3)-dimensional NCYM.

The (1 + 3)-dimensional open D-string found in section 3 resolves this
puzzle. First the existence of this theory is consistent with the
S-duality. Second that this theory is equivalent to the usual 
(1 + 3)-dimensional NCYM as discussed in section 3 is also consistent
with the S-duality between the (1 + 3)-dimensional NCOS and the usual
NCYM. Our interpretation for S-duality is a bit different from that
given in~\cite{gopmms} where a worldvolume Poincare duality is also
employed as discussed in section 4. 
In terms of our interpretation,  the 
(1 + 3)-dimensional NCYM is actually S-dual invariant while our open
D-string theory is S-dual to the NCOS.

	Our picture of S-duality for the decoupled theories from the
parent type IIB string theory is as follows: In general, a decoupled
open brane theory is S-dual to another decoupled open brane theory 
while a decoupled field theory is S-dual to another decoupled field
theory. The examples are: a) (1 + 3)-dimensional NCOS is S-dual to the
(1 + 3)-dimensional open D-string theory in this paper, 
(1 + 5)-dimensional NCOS is S-dual to the (1 + 5)-dimensional OD1 theory 
and (1 + 5)-dimensional
OD3 is S-dual to the (1 + 5)-dimensional open D3 brane theory in this
paper. The usual (1 + 3)-dimensional NCYM is S-dual to the 
(1 + 3)-dimensional NCYM discussed in section 4 (actually self-dual),
 the usual (1 + 5)-dimensional NCYM is S-dual to the (1 + 5)-dimensional NCYM
discussed in the previous section. As discussed in the previous section,
an open brane theory should not be in general S-dual to a field theory
since the latter may not be complete (due to nonrenomalizability) 
while the former is generally  complete. 

	As mentioned above, the reason that the usual (1 +
3)-dimensional NCYM can be S-dual (using the interpretation of
~\cite{gopmms})  to the (1 + 3)-dimensional NCOS is
due to that this NCYM is a complete theory and is equivalent to 
the (1 + 3)-dimensional open D-string theory. This is, however, not the
case in (1 + 5)-dimensions. 

	Now the remaining question is: Does a general (1 +
3)-dimensional (p, q) open string theory exist? The answer should be yes
if the type IIB SL(2, Z) is inherited to the little type IIb string
theory. The existences of both (1 + 3)-dimensional NCOS and open D-string
theories, both (1 + 5)-dimensional NCOS and OD1 and the two versions of
open D3 brane theory related by S-duality also strongly support
this. Given that an open (p, q)-string can end on D3 branes, one expects
that a force due to a proper background field can balance the
tension. For examples, in the simplest context, if we apply only a
near-critical electric field $B_{01}$, 
\begin{equation}
2 \pi \a' \e^{01} B_{01} = 1 - \frac{\e}{2},
\end{equation}
with the usual scaling limit for NCOS,
\begin{equation}
\a' = \e \a'_{\rm eff}, \ \ g_s = \frac{G^2_o}{\sqrt \e},
\end{equation}
we have 
\begin{equation}
- \frac{1}{2\pi \a'} \sqrt{p^2 + q^2/g^2_s} + p \e^{01} B_{01}
= - \frac{1}{4\pi p \a'_{\rm eff}} (p^2 + \frac{q^2}{G^4_{o}})
\end{equation}
which is finite and is the tension for the decoupled theory which is
still a NCOS. Similarly, we can have only a near-critical RR $C_{01}$ and
with the decoupling limit for the open D-string theory, we can also end
up with a deformed open D-string theory. 

Recall that an open  
(p, q)-string is a non-threshold bound and its ends carry both NSNS and
RR charges (or electric and magnetic charges with respect to the D3
brane worldvolume gauge field). So both the background NSNS $B_{01}$ and 
$C_{01}$ apply forces on this string. 
A genuine open (p, q)-string theory requires the presence of both the 
near-critical field $B_{01}$ and $C_{01}$. Further each of these two fields
along with the proper scalings for the closed string coupling and the bulk
metric must act in a non-trivial way such that we can end up with
 a finite tension for the (p, q)-string theory.       
One can check easily that a naive critical field limit following that 
for either open D-string theory or NCOS does not work. The investigation on
this is in progress and we hope to report this elsewhere. Nevertheless,
the finite tension is expected to be
\begin{equation}
T_{(p, q)} = \frac{1}{2\pi\a'_{\rm eff}} \sqrt{p^2 + \frac{q^2}{G^4_o}}.
\end{equation}
This should also be true for the (1 + 5)-dimensional open (p, q)-string
theory. We expect that the (p, q)-string action proposed 
in~\cite{towtwo,towthree} may be useful. 

\bigskip

\noindent
{\it Note Added}: During the course of writing up, we become aware that 
when the spatial directions of the D(p - 2) brane are compactified,
our (1 + p)-dimensional open D(p - 2) brane theory may be related to 
the Galilean D(p - 2) brane theory discovered 
in ~\cite{gomo} (see also~\cite{danone,dantwo}). However, there are
 differences between these two theories. Let us mention a few: 
1)The spatial directions of the brane for our 
open D(p - 2) brane theory can be either non-compact or compact while 
by definition the spatial directions of the
brane for the Galilean D(p - 2) brane
theory found in~\cite{gomo,danone} must be compact due to the absence of the
base D-brane. 2) As a result, our open D(p - 2) brane theory lives on (1
+ p)-dimensional Dp brane worldvolume while the Galilean D(p - 2) brane
theory lives on the (1 + 9)-dimensional spacetime. 3) The starting
points are completely different.

\section*{Acknowledgements}
The author would like to thank R.G. Cai for bringing his attention to
reference~\cite{ber}, to thank Mike Duff, Hong Lu and Yong-Shi Wu 
for discussions.
The author would also like to thank J. P. van der Schaar for discussion
on the open (p, q)-string theory and
acknowledges the support of U. S. Department of Energy.


\end{document}